\documentclass[11pt,a4paper]{article}
\usepackage[T1]{fontenc}
\usepackage{float}     
\usepackage{geometry}     
\usepackage{graphicx,epsf,subfigure}
\usepackage{indentfirst}
\usepackage{color}
\usepackage{setspace}
\author{
Yuko Aoyanagi\textit{$^{a,b}$} \and J\'er\'emy Hure$^{\ast}$\textit{$^{b}$} \and Jos\'e Bico\textit{$^{b}$}  \and Beno\^it Roman\textit{$^{b}$}
}

\title{Random blisters on stickers: metrology through defects}

\begin{document}

\maketitle

\begin{abstract}
Blisters are commonly observed when an adhesive sheet is carelessly deposited on a plate. 
Although such blisters are usually not desired for practical applications, we show through model 
experiments on angular blisters how material properties can be deduced from height profile 
measurements. 
In particular the typical curvature of the crests is found proportional to an {\it elasto-capillary} length that compares the bending stiffness of the sheet with adhesion energy. 
In addition, the radius of the tip allows to estimate the product of this length with the thickness of the sheet.
The relevance of these results to realistic random configurations is finally confirmed.

\end{abstract}

\section{Introduction}

\footnotetext{\textit{$^{a}$~Dept. of Physics, Ochanomizu University, 2-1-1 Otsuka, Bunkyo-ku, Tokyo 112-8610, Japan.}}
\footnotetext{\textit{$^{b}$~Physique et M\'ecanique des Milieux H\'et\'erog\`enes, CNRS UMR 7636, UPMC \& Univ. Paris Diderot, ESPCI-ParisTech, 10 rue Vauquelin, 75231 Paris Cedex 05, France. E-mail: jeremy.hure@espci.fr.}}

Placing a sticker on a flat surface is usually a delicate task since blisters form very easily.
Air may indeed remain trapped between the adhesive sheet and the substrate leading to bubble-shaped blisters, while linear folds and wrinkled structures may appear when mismatching 
adhesion fronts are brought together. 
The consequences of blisters are important in a wide range of industrial applications involving the 
deposition of an adhesive sheet on a panel where such defects are crucially undesired.
Similar blisters are also observed at  nanoscale when graphene sheets are deposited on 
substrates  \cite{schniepp08, li09, xu09, zong10}. 
Beyond deposition processes, blisters are also commonly observed when materials covered with  
thin films delaminate. Indeed  differential thermal expansion, differential swelling or residual stresses due the coating 
process can lead to the formation of wrinkles \cite{bowden98} and eventually to the debonding of 
the film if these stresses are compressive \cite{hutchinson92, gioia97, hutchinson99, faulhaber06, mei2007}. While 
most studies have focused on predicting the onset of buckling with the aim of 
avoiding disastrous defects in thin film coatings,
 recent experiments have 
shown how surfaces can be micro-patterned  with controlled wrinkles and blisters \cite
{bowden98, breid09}, with potential applications spanning from optics \cite{chan06} to stretchable electronics \cite {ko2008, vella2009, rogers10}. 
Another interesting application of wrinkles and blisters, following studies on the {\it blister test}  \cite{dannenberg61, williams69}, is to use their geometrical characteristics to probe the mechanical properties of thin sheets as well as the energy of adhesion \cite{stafford2004, huang07, khang09}.
Motivated by these novel metrology techniques, we propose  to study which relevant 
information may be deduced from monitoring the shapes of the blisters obtained when a thin adhesive 
sheet is randomly deposited on a flat substrate.
We first describe a model experiment complementing, in a different regime, the work from Chopin {\it 
et al.} \cite{chopin2008}. In this experiment adhesion is due to a film of wetting liquid preliminary coated on the substrate and a 
single angular blister is produced by compressing one edge of the sheet. Results from numerical 
simulations are in addition compared with experimental data.
In a second section, the outputs of the simplified configuration are applied to complex blisters 
obtained by depositing carelessly a thin sheet on an adhesive (non slippery) substrate. The different physical parameters deduced from monitoring the curvature of the crest and the radius of the tip of the blisters are finally compared with their estimates from other techniques. 

\section{Model experiment with a single blister}

Consider an adhesive film carelessly thrown over a substrate.
The parts of the falling adhesive sheet that first touch the substrate immediately stick to it. Around these contact regions, adhesion zones develop and  are  generally not compatible with the other patches, which leads to the formation of  blisters.
We usually observe  
that the blisters often lie along a straight direction, with a slowly varying width and a finite length $L.$
 In this section we study a model experiment on isolated {\it angular} blisters with such properties.

\subsection{Experimental setup and  typical observations }

\begin{figure}[H]
\begin{minipage}[b]{0.5\textwidth}
\subfigure[]{\includegraphics[width=8.5cm]{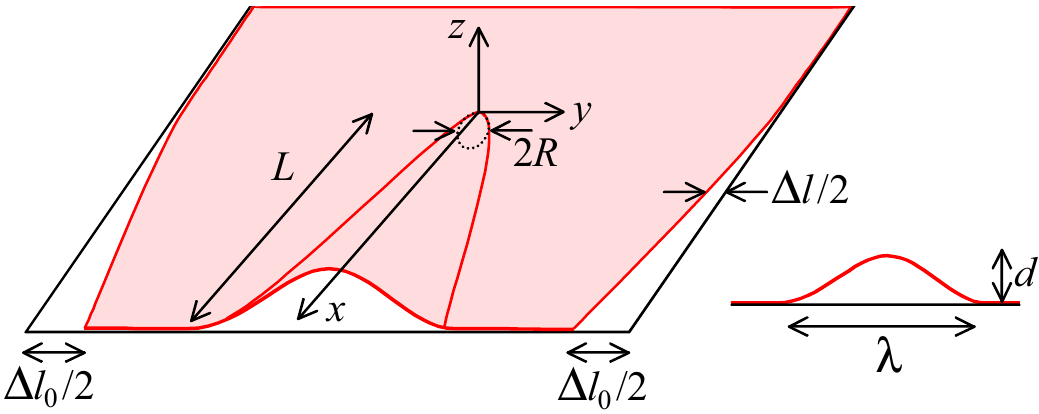}}
\end{minipage}
\begin{minipage}[b]{0.5\textwidth}
\subfigure[]{\includegraphics[width=8.5cm]{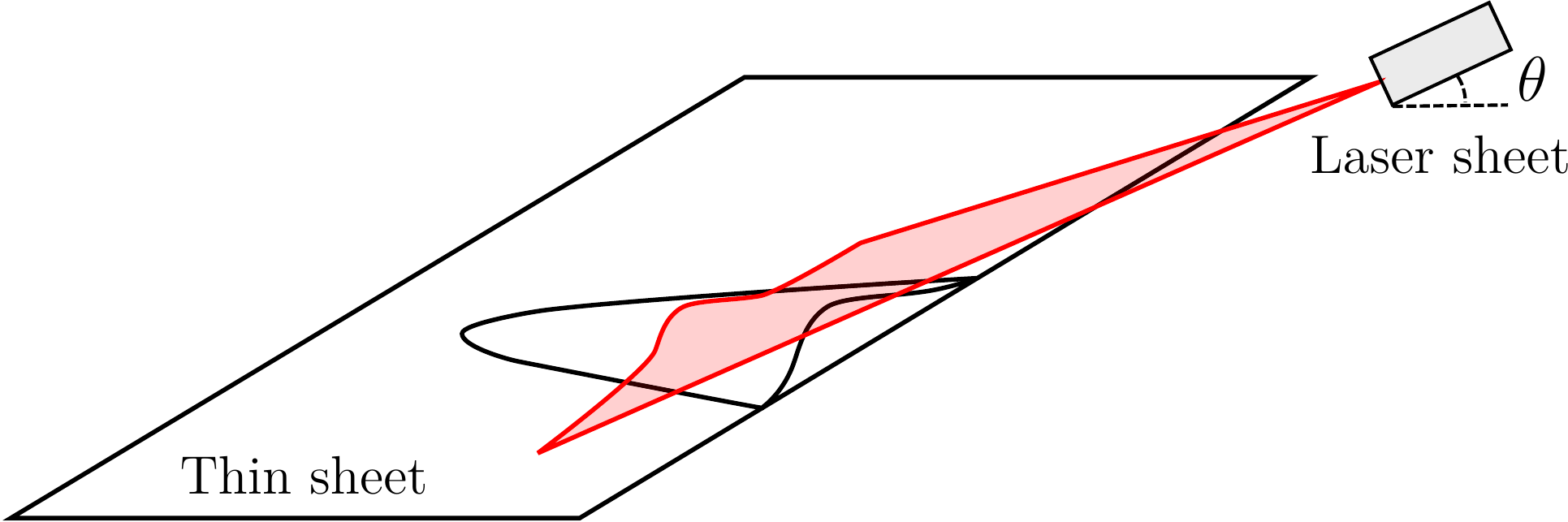}}
\end{minipage}
\caption{(a) Experimental setup: a  thin sheet is deposited over a rigid plate coated with ethanol. A 
transverse displacement $\Delta l_0$ is imposed at one edge with a caliper symbolized by the arrows, leading to the formation of a single blister. The shape of the blister is defined by its local height $d$ and width $\lambda$ associated with a local transverse displacement $\Delta l$. (b) A laser sheet tilted by an angle $\theta$ is used to measure the height profile of the blister.}
\label{fig:drawing}
\end{figure}

Thin sheets of biaxially oriented polypropylene (Innovia films) are laid down a rigid plate covered 
with a layer of ethanol. The Young's modulus and Poisson's ratio of the polymer are $E = 2.6\pm0.2\,
$GPa$^1$ \footnote{$^1$ The range given for Young's modulus corresponds to the anisotropy due to the process of fabrication.} and $\nu = 0.4$, respectively. Three film thicknesses were selected for our experiments: $h
$=15, 30 and 50$\,\mu$m.
Due to the surface tension of the liquid ($\gamma=22.4\,\mathrm{mN.m^{-1}}$), the sheets adhere 
to the plate (sheets and plate are totally wetted by ethanol). 
As sketched in figure~\ref{fig:drawing}a, a controlled transverse displacement is imposed at one edge, 
leading to the formation of a single blister that propagates through the sheet up to a finite distance $L$ from the compressed edge. 
The shape of the observed structures evolves slightly with the thickness of the liquid layer, but 
a quasi-steady regime is achieved as the liquid progressively evaporates. 
Our measurements are conducted in this regime, where the shape is independent of the remaining volume of liquid.

\begin{figure}[!h]
\begin{minipage}[b]{0.5\textwidth}
\subfigure[]{\includegraphics[width=7.5cm]{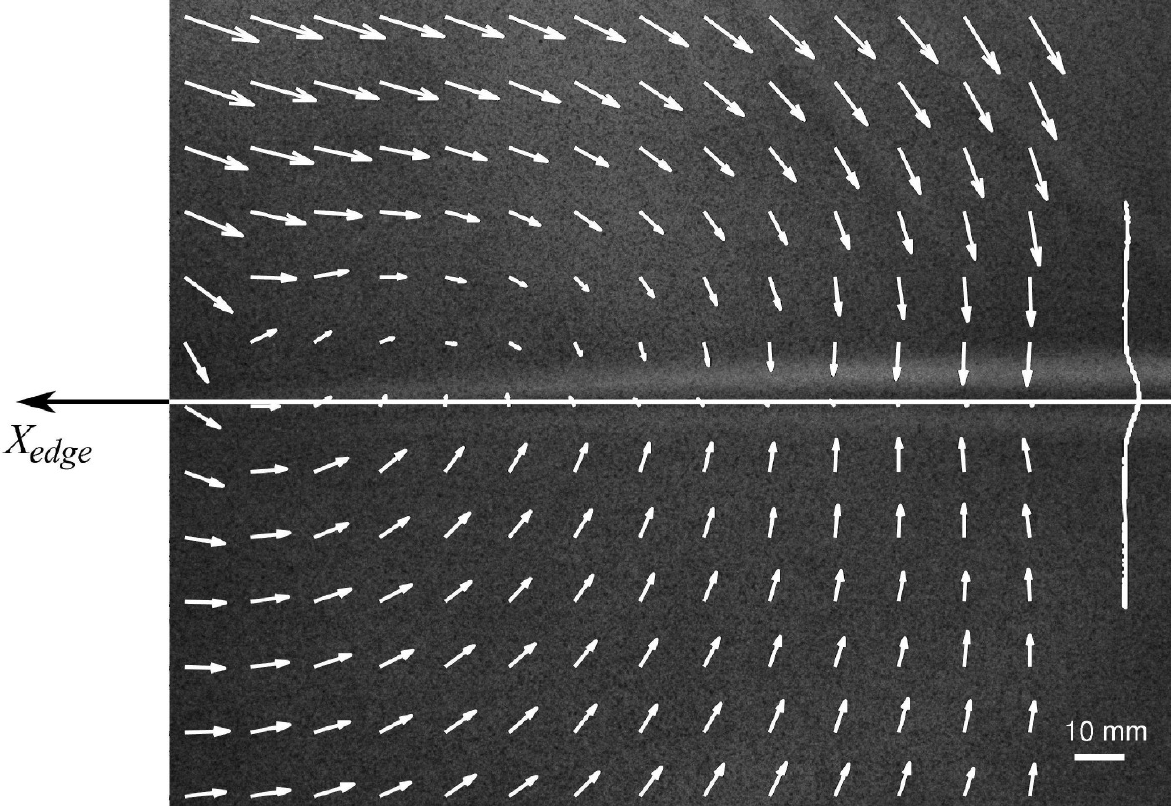}}
\end{minipage}
\begin{minipage}[b]{0.5\textwidth}
\subfigure[]{\includegraphics[width=7.5cm]{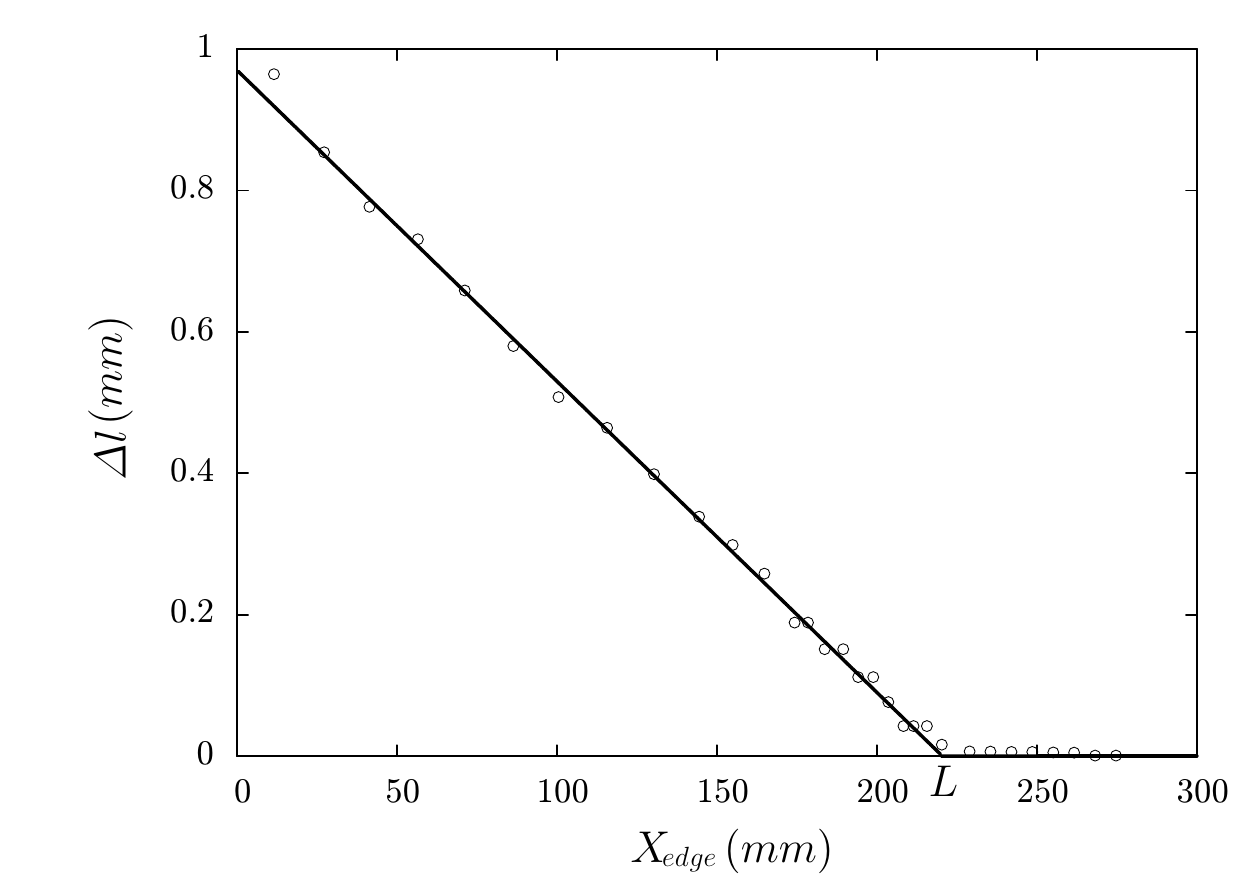}}
\end{minipage}
\begin{minipage}[b]{0.5\textwidth}
\subfigure[]{\includegraphics[width=7.5cm]{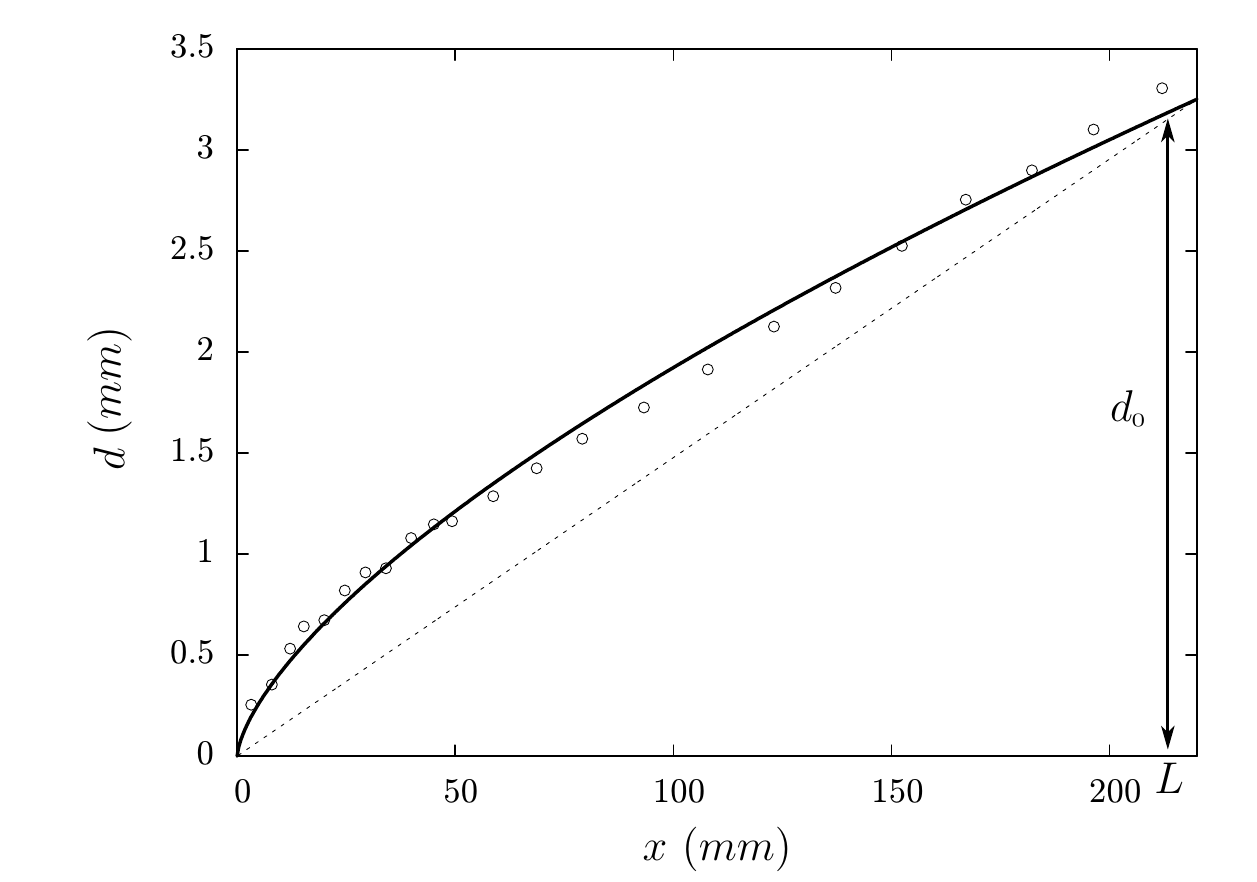}}
\end{minipage}
\begin{minipage}[b]{0.5\textwidth}
\subfigure[]{\includegraphics[width=7.5cm]{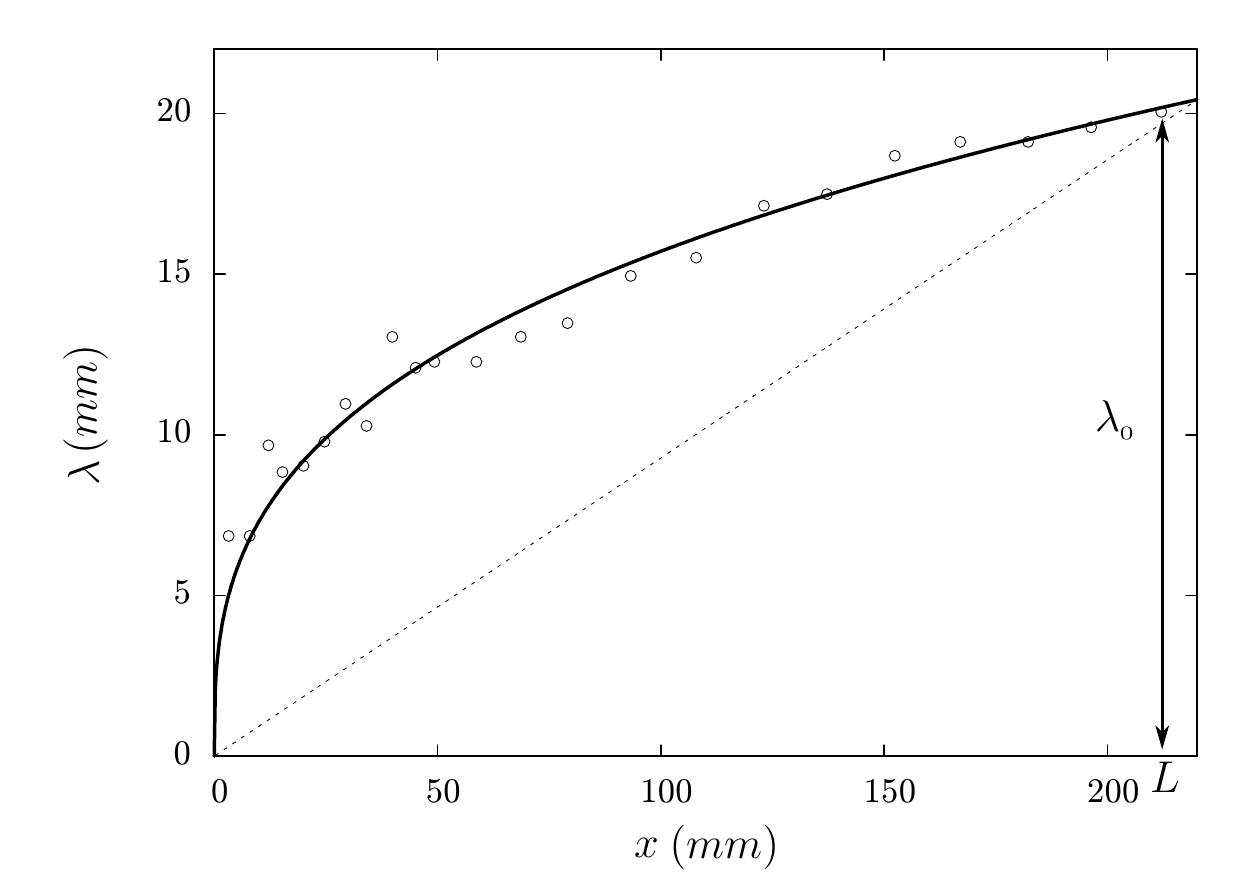}}
\end{minipage}
\caption{Typical experimental procedure ($h=30\,\mu$m, $\Delta l_0=0.97\,$mm). (a) The profile of 
the blister is determined by scanning the blister with a tilted laser sheet, while a standard image 
correlation technique is used to estimate the local displacement along the blister: the white line on the right corresponds to the intersection of the laser sheet with the thin sheet; the white arrows represent the in-plane displacement field. (b) The 
transverse displacement along the blister is found to vary linearly with the distance to
the edge $X_{edge}$, which allows for a precise definition of the 
length of the blister $L$ and the position of its tip.  (c,d) Height and width as a function of the distance from the tip $x=L-X_{edge}$. The experimental data are fitted by power functions with respectively $2/3$ and $1/3$ 
exponents (dashed lines: conical situation).}
\label{fig2} 
\end{figure}

The planar deformation of the sheet is measured using a standard image correlation 
technique (DaVis software from LaVision): the thin films are sputtered with black paint and a picture 
taken after displacement is compared with a reference image of the non-deformed film. The cross-correlation of the two pictures leads to the planar displacement field 
of the sheet (arrows  in figure \ref{fig2}a).
 Comparing the motion of the sheet on both sides of the blister gives a measurement of the transverse displacement $\Delta l$ along the blister as a function of the distance to the edge $X_{edge}$. We observe a linear variation of this transverse displacement with $X_{edge}$ (figure~\ref
{fig2}b), which shows that the opposite sides of the thin sheet separated by the blister undergo a relative rotation motion (see also arrows in figure~\ref{fig2}a) around the tip of the blister.
Indeed, both patches move as rigid bodies (a composition of a translation and a rotation) with respect to each other. 
At the tip of the blister, this relative motion is zero ($\Delta l=0$), so that both patches rotate around it.
Since the localization of the center of rotation is very accurate from these measurements, we define the length $L$ of the blister as the distance of this center to the edge. In the following of the paper we choose this center as the origin of our coordinate system (with $x = L-X_{edge}$).

In our experiments, the length $L$ of the blister was not reproducible, and we do not give a prediction for this parameter in this paper. Indeed, the propagation of the blister is stopped by secondary wrinkles and blisters forming in front of the tip.
As a matter of fact, we expect the rotational motion of the thin sheet to generate ortho-radial tension ahead of the tip, inducing radial compression (due to Poisson's ratio), which in turn leads to the formation of wrinkles or secondary  blisters perpendicular to the main angular blister. These smaller structures are observed in the vicinity of  the tip and are more numerous for thinner sheets$^2$. \footnote{$^2$
Such wrinkles are reminiscent of secondary cracks than can appear due to tensile stresses parallel  to a main crack at its tip in the presence of a locally weak interface. They act  as crack-stoppers, as described in the Cook-Gordon mechanism \cite{gordon}.} We focus now on the profile of the blister along its axis.
The width $\lambda$, the amplitude $d$ and the transverse displacement $\Delta l$ are measured 
as a function of the distance to the tip along the axis $x,$ for a given displacement $\Delta l_0$ at the 
edge. The height profile is obtained by scanning a laser sheet tilted by an angle $\theta$ along the blister (see figure \ref{fig:drawing}b). After calibration, the in-plane displacement of the laser sheet seen from above gives the out-of-plane displacement \cite{vella2009}. Thus, the shiny line 
observed at the intersection of the laser sheet with the blister directly gives the local profile (figure~\ref
{fig2}a). The amplitude and width are measured by fitting the profile by a cosine function $z(y) = d/2[1+\cos(2 \pi y /\lambda)]$, since this form is classically observed with beams slightly bent.
The variations of $d(x)$ and $\lambda(x)$ are respectively displayed in figures~\ref{fig2}c and~\ref
{fig2}d.

\begin{figure}[!h]
\centering
\includegraphics[width=10cm]{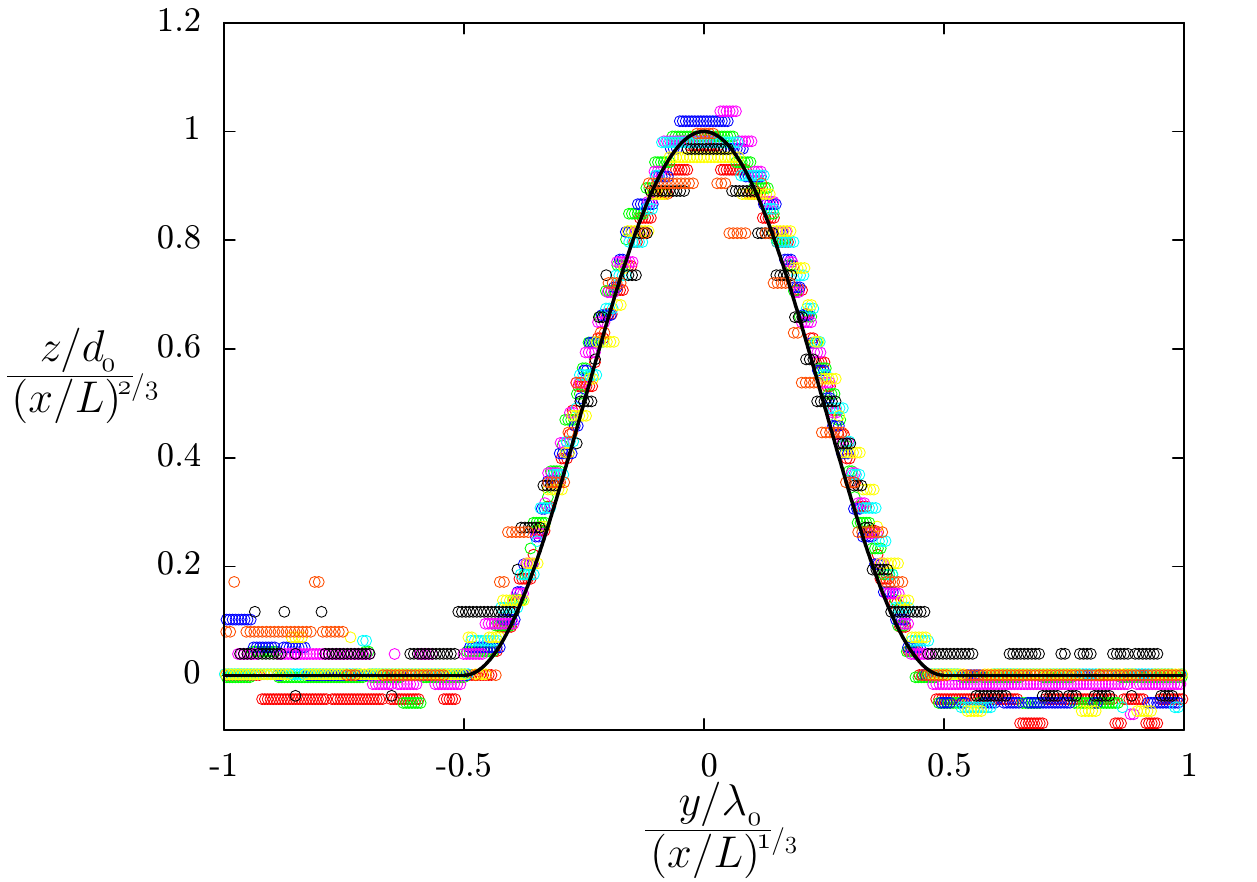}
\caption{Profiles of the successive sections of a same blister represented in the normalized 
coordinates $(z/d_0)/(x/L)^{2/3}$ {\it vs.} $(y/\lambda_0)/(x/L)^{1/3}$ (Equations (\ref{eq})). Solid line corresponds to $1/2[1+\cos(2\pi x)]$.}
\label{rescale}
\end{figure}

Plotting the profiles of different sections of the same blister in {\it stretched} coordinates, with $z$ and 
$y$ respectively normalized by $x^{2/3}$ and $x^{1/3}$, reveals a self-similar shape (figure~\ref
{rescale}).
However, contrary to the case of the {\it liquid blister test} studied by Chopin {\it et al.} \cite
{chopin2008}, the profile is not conical (which would correspond to both $\lambda$ and $d$ being
proportional to the distance to the tip $x$).
The aim of the following section is to rationalize these observations.

\subsection{1D {\it vs.} 2D blisters}

We first recall the case of the {\it one-dimensional} blisters  \cite{kendall76}, where a buckled elastic line 
adheres on a substrate. A balance between adhesion (which favors large areas of contact)
and bending (which is minimized for blisters of large widths) energies
 dictates the shape of the observed blisters.
Indeed, within the limit of small amplitudes, the adhesion energy density scales as $\mathcal{E}_\gamma\sim \gamma$,  while the bending energy density is proportional to $\mathcal{E}_b \sim B(d/\lambda^2)^2$, where 
$B=Eh^3/12(1-\nu^2)$ is the bending modulus of the film. At equilibrium, both energies are of same order, which leads to a constant 
curvature of the crest:
\begin{equation}
\lambda^2/d \sim \sqrt{B/\gamma} =  L_{ec}. 
\label{eqelcap}
\end{equation}
The characteristic length scale $L_{ec}$ compares surface tension (or adhesion) to bending 
stiffness and has been refereed  to as {\it elasto-capillary} length \cite{cohen2003,bico04,review10}. 

\begin{figure}[!h]
\centering
\includegraphics[width=10cm]{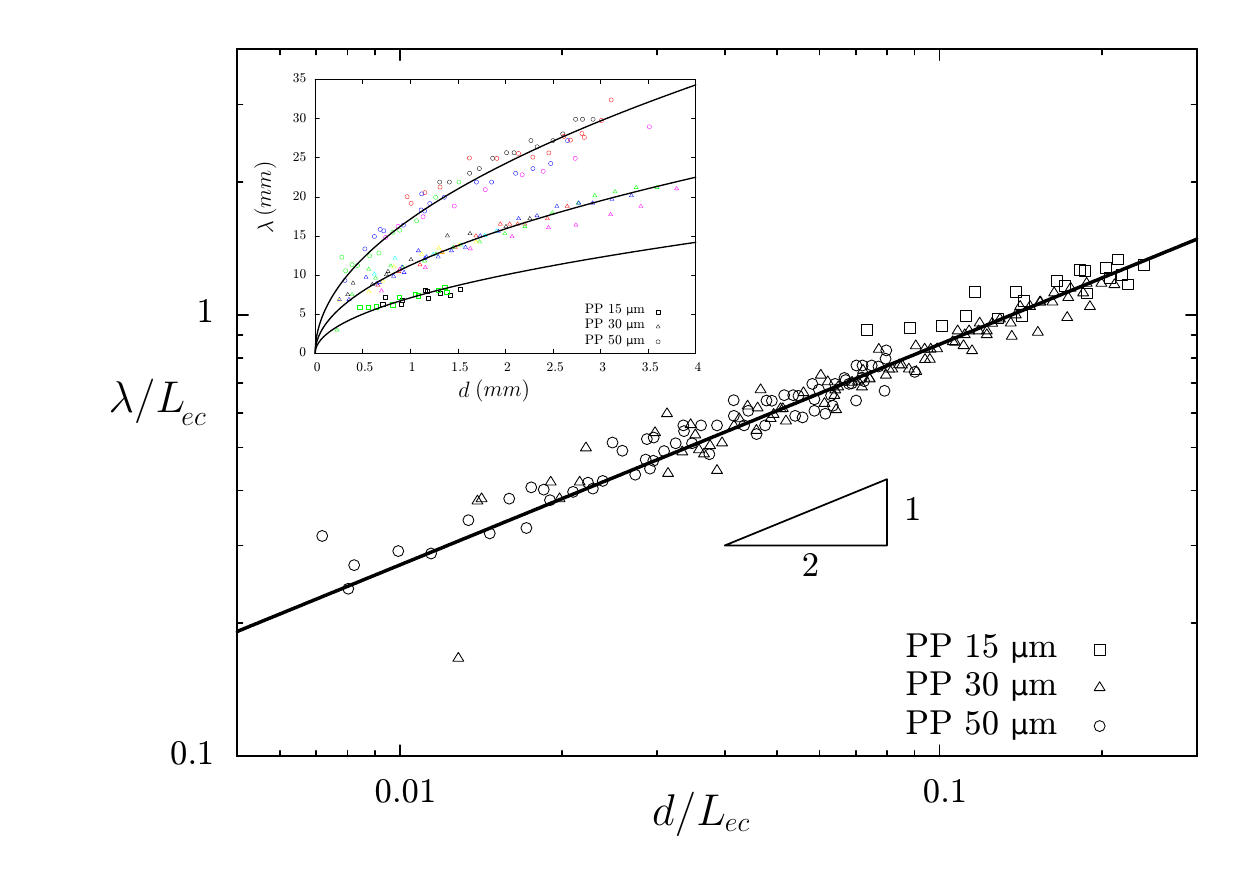}
\caption{Normalized width $\lambda/L_{ec}$ as a function of normalized amplitude $d/L_{ec}$ for 
experiments conducted with samples of different thicknesses and initial transverse displacements. The solid line corresponds to a power law fit with an exponent 1/2: $\lambda^2/d = 7.3 L_{ec}$ (see Equations (\ref{eqelcap}) and (\ref{eqLecpref})). Inset: 
dimensional data. Colored symbols are used to differentiate the different experiments conducted with sheets of the same thickness.}
\label{fig3}
\end{figure}

We compare this result with our 2D experiments conducted with samples of different thicknesses $h$ and initial transverse displacements $
\Delta l_0$  in figure~\ref{fig3}. The width $\lambda$ is  found proportional to $d^{1/2}, $ with a different prefactor depending only on the thickness $h$ (inset).
Moreover, all the experimental data are found to collapse into a single power law $\lambda \sim \sqrt{d L_{ec}}.$ 
This result suggests that all the scanned profiles of the blisters behave as if they were isolated 1D blisters.

\begin{figure}[!h]
\begin{minipage}[b]{0.5\textwidth}
\subfigure[]{\includegraphics[width=7.5cm]{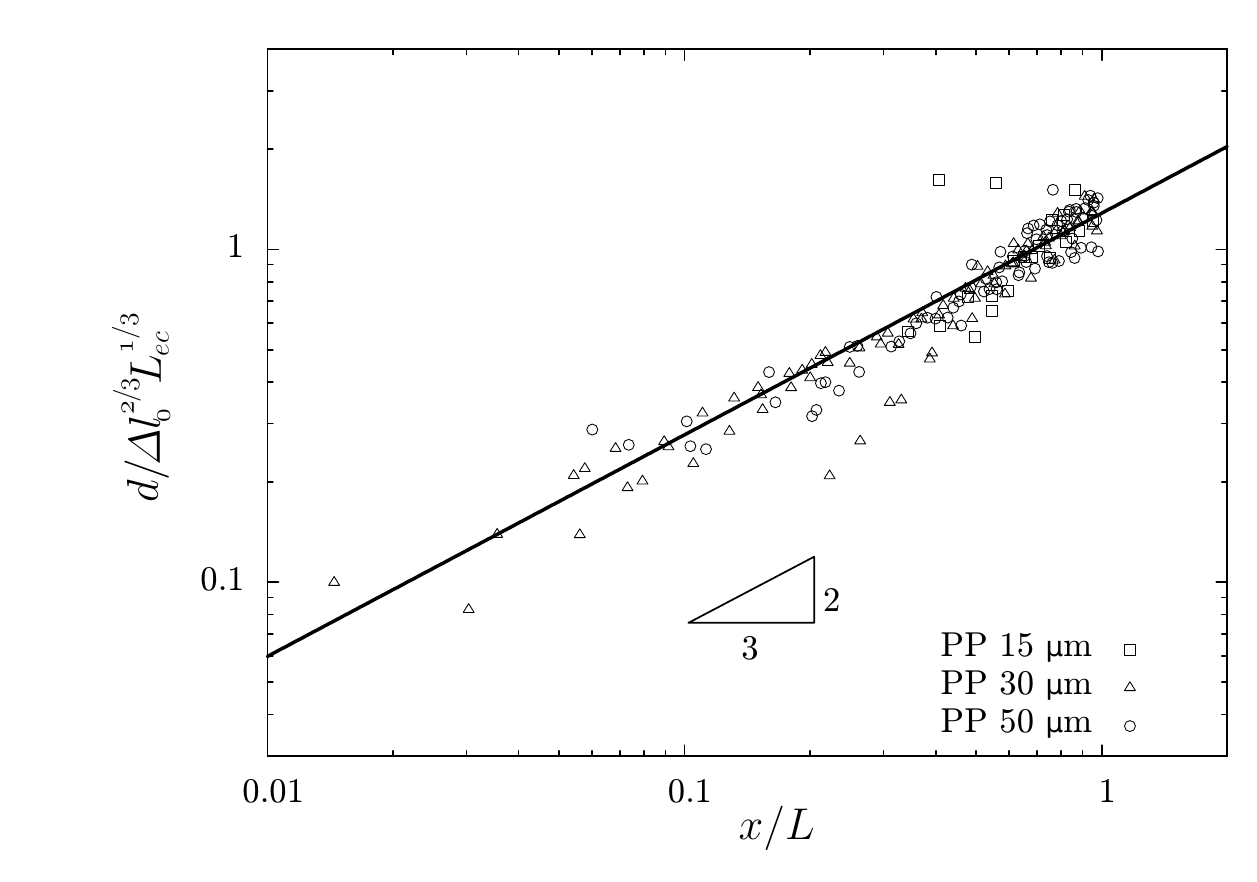}}
\end{minipage}
\begin{minipage}[b]{0.5\textwidth}
\subfigure[]{\includegraphics[width=7.5cm]{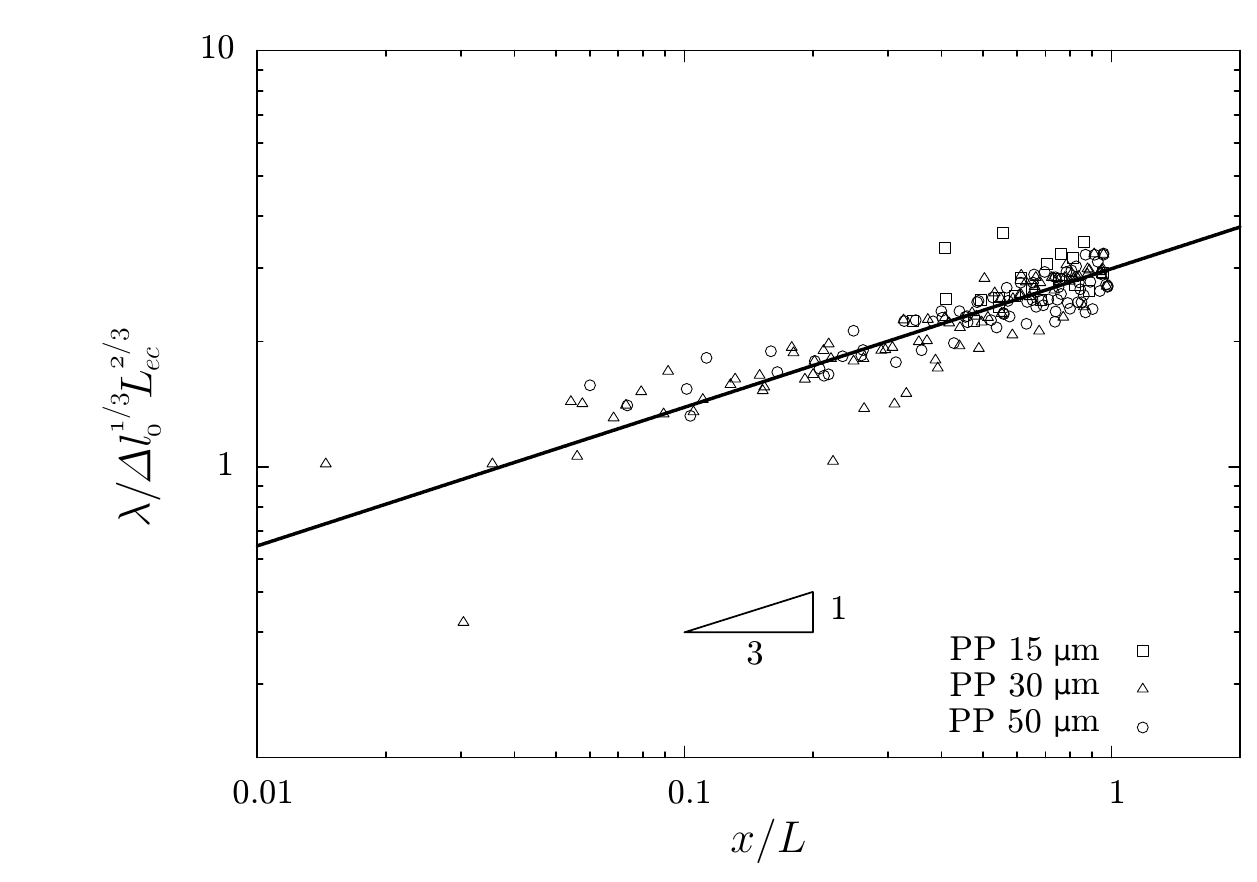}}
\end{minipage}
\caption{(a,b) Rescaled height and width as a function of the dimensionless distance from the tip (Equations (\ref{eq})). Solid lines correspond to power law fits with respective exponents 2/3 and 1/3.}
\label{fig4}
\end{figure}

\noindent
We have seen that the transverse displacement follows a linear law: 
\begin{equation}
\Delta l = \Delta l_0 x/L.
\end{equation}
Along a cross section of the blister  conservation of length imposes that the total arc length of the 
blister is equal to the sum $\lambda(x) + \Delta l(x)$, which leads (in the limit of small amplitudes) 
to:
\begin{equation}
d(x)^2 \sim \lambda(x)\Delta l(x).
\end{equation}
The combination of this last geometrical relation with the expressions for the transverse displacement and 
the curvature finally gives the variations of the amplitude and the  width of the blister with the distance from the tip:
\begin{equation}
d/\Delta l_0^{2/3}L_{ec}^{1/3} \sim  \left(x/L\right)^{2/3}  ~~~~~~\mathrm{and} ~~~~~~\lambda/\Delta l_0^{1/3}L_{ec}^{2/3} \sim \left(x/L\right)^{1/3}, 
\label{eq}
\end{equation}
which is in agreement with the self-similar profiles displayed in figure~\ref{rescale}.
The data corresponding to samples of different thicknesses and initial transverse displacements are 
superposed in figures~\ref{fig4}a and \ref{fig4}b. 
Although scattered, the data collapse into the predicted master curves over two orders of 
magnitude (measurements obtained close to the tip tend to be noisy due to the very small 
amplitude involved and to the presence of secondary wrinkles in the vicinity of the tip).  

\subsection{Stretching energy in 2D blisters}

We have seen that the shape of a blister can be viewed as a continuous succession of profiles corresponding to one-dimensional blisters. However, this shape should correspond to the minimum of the global energy of the system.
Contrary to the case of an elastic line,  the bending energy of a plate involves the two principal curvatures. However, if the curvature of the crest line is not too high ($d''(x) \ll  d/\lambda^2$), the bending energy of the blister reduces to the 1D estimate. 

\begin{figure}[!h]
\begin{minipage}[b]{0.5\textwidth}
\subfigure[]{\includegraphics[width=7.5cm]{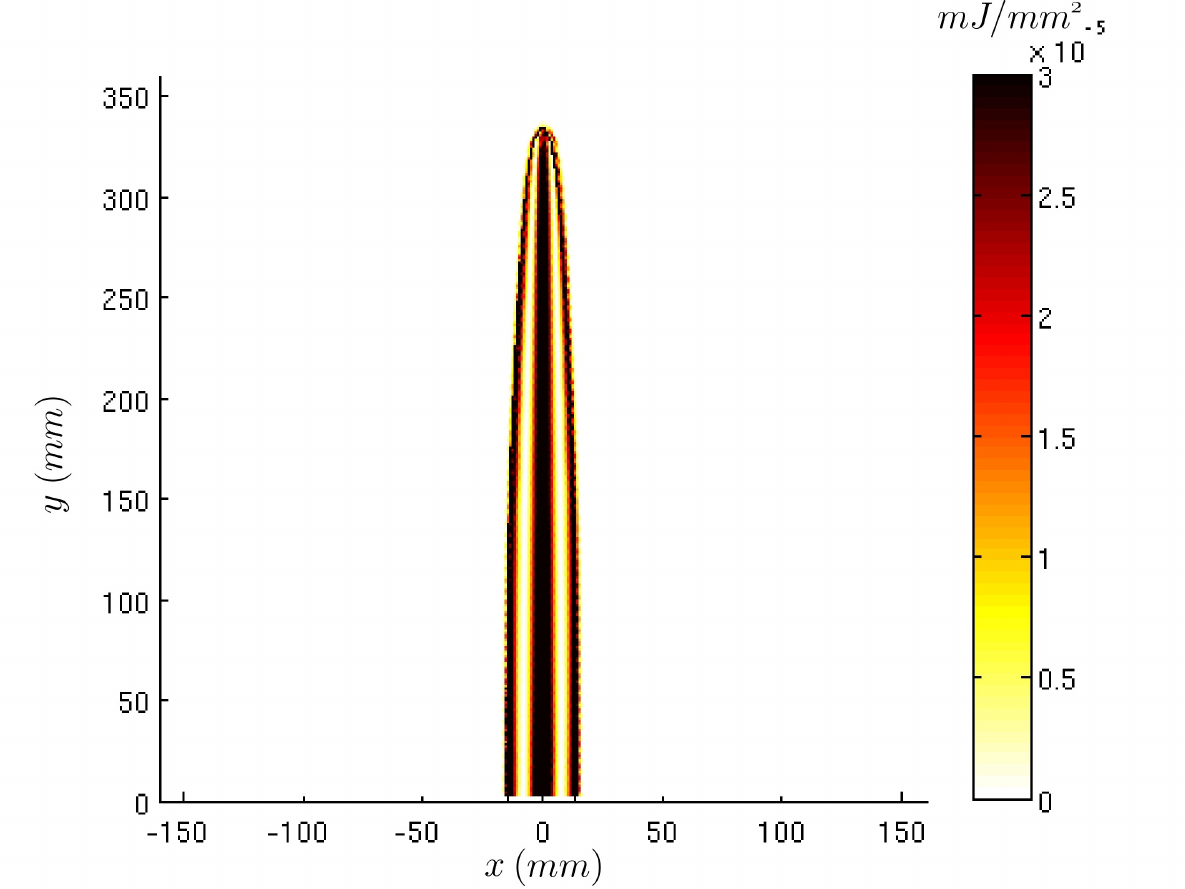}}
\end{minipage}
\begin{minipage}[b]{0.5\textwidth}
\subfigure[]{\includegraphics[width=7.5cm]{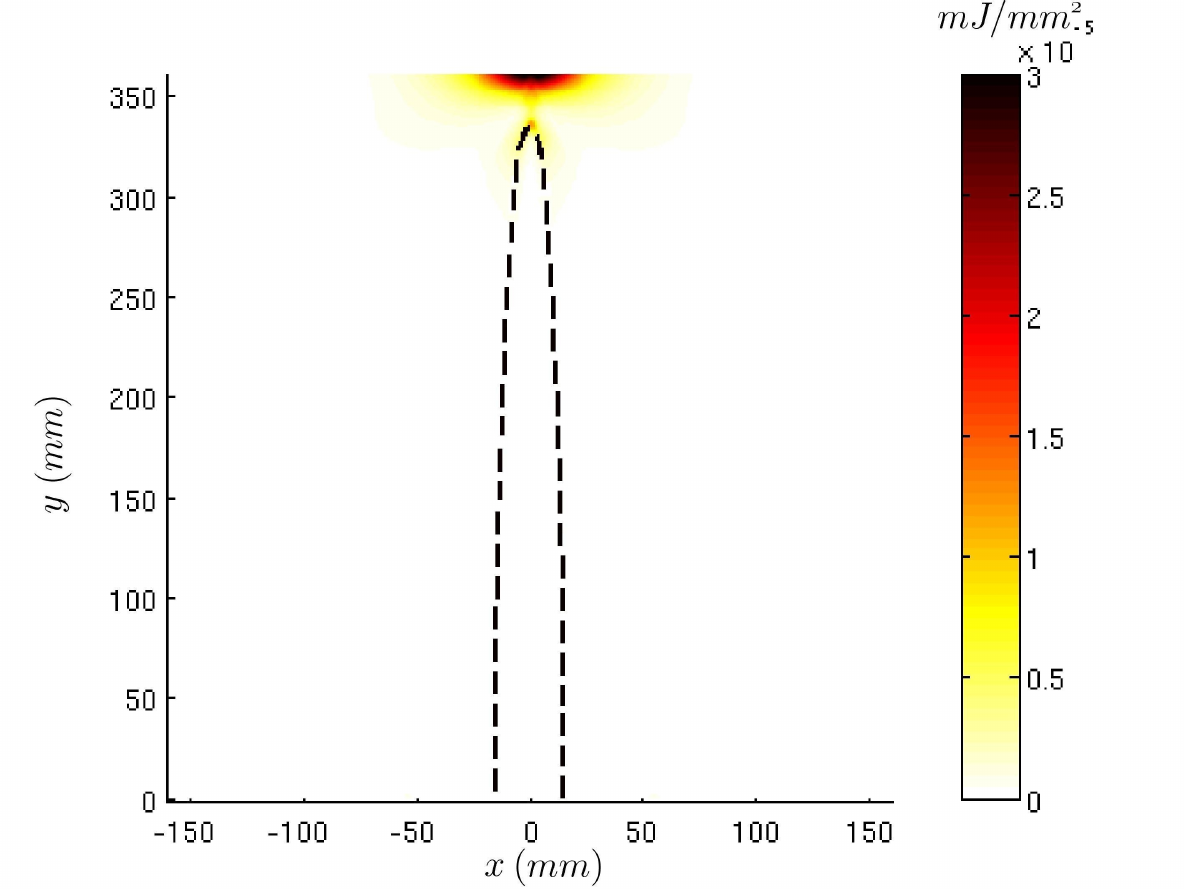}}
\end{minipage}
\caption{Numerical simulation using the Surface Evolver software: $E=2000\,$MPa, $\nu~=~0.4$, $h=50
\,\mu$m, $\Delta l_0=1\,$mm and $\gamma=22.4\,\mathrm{mN.m^{-1}}$. Maps of bending (a) and 
stretching (b) energy densities confirm that the stretching energy inside the blister is negligible (dashed line: blister's edge).}
\label{fig6}
\end{figure}

More importantly, since the thin sheet is curved along the crest of the blister  in its two principal directions, it does not conserve its initial zero gaussian curvature (the product of both principal curvatures). According to Gauss' \textit{Theorema Egregium} \cite{struik88}, this change in gaussian curvature implies stretching or compression in the thin sheet. Stretching energy is usually so dominant in thin objects, that it leads to the formation of crumpling singularities to minimize it \cite{witten07}. One way to estimate the stretching energy is to compare the actual arc length of the crest  to the length of the non stretched  ideal cone (figure \ref{fig2}c). If the profile is given by $d(x)$, the curvilinear length is obtained from the integral $\int_0^L (1 + 
d'(x)^2)^{1/2} dx $, while the non stretched length is simply $ L(1 + (d_0/L)^2)^{1/2}$. If $d(x)$ follows a power law, the typical strain is thus given by $\epsilon \sim 
(d_0/L)^2$.
We finally obtain an estimate for the stretching energy density inside the blister: $\mathcal{E}_s \sim Eh (d_0/L)^4$. 
We expect the description of the blister as a succession of 1D profiles to be valid in the limit where $\mathcal{E}_s$ is negligible when compared to $\mathcal{E}_
\gamma$ (or $\mathcal{E}_b$), which corresponds to:
\begin{equation}
 d_0 \ll \left(\frac{\gamma}{Eh}\right)^{1/4} L. 
 \label{eqnostretch}
\end{equation}
In our experiments both $( \gamma/Eh)^{1/4}$ and $d_0/L$ are typically on the order of  2\%. Nevertheless all numerical prefactors have been omitted in our description, and the stretching energy may have been largely overestimated. 
In order to check the validity of these estimates, numerical simulations were carried out with the Surface Evolver software \cite{brakke92}.
The physical parameters selected for the simulation correspond to a 
typical experiment: $E=2000\,$MPa, $\nu$=0.4, $h=50\,\mu$m, $\Delta l_0=1\,$mm and $
\gamma=22.4\,\mathrm{mN.m^{-1}}$ (corresponding to $( \gamma/Eh)^{1/4} \simeq 2\%$). The plate is discretized into finite elements with an elastic energy corresponding to their deformation. Adhesion energy is represented by a potential decaying exponentially with the distance from the rigid plate (we checked that the characteristic decay length of the potential does not influence the final result, as long as it remains small). Surface Evolver minimizes the energy of the system using alternately gradient and conjugate gradient algorithms. Stochasticity is introduced by ``jiggling'' the position of vertices to escape from local minima of energy. Mappings of the bending and stretching energy densities are displayed in figures~\ref{fig6}a and \ref{fig6}b. 
The maximal values in the blister, respectively  $4.10^{-5}\,\mathrm{mJ.mm^{-2}}$ and $3.10^{-7}\,\mathrm{mJ.mm^{-2}}$ for bending and stretching energy densities, are consistent with the scalings $B/L_{ec}^2 \sim 10^{-5}\,\mathrm{mJ.mm^{-2}}$ and $Eh(d_0/L)^4 \sim 10^{-6}\,\mathrm{mJ.mm^{-2}}$. These results confirm that stretching energy along  the blister is indeed negligible in this regime. We also checked that the numerical profiles obeyed equations (\ref{eq}).

In the opposite case, where (\ref{eqnostretch}) is not satisfied, stretching energy plays a dominant role, and the previous scalings (\ref{eq}) are no longer valid. As a consequence, the shape of the blister that minimizes the energy of the system is a developable cone. This 
limit is discussed in details by Chopin {\it et al.} \cite{chopin2008}. Indeed, a typical value for $d_0$ in their {\it liquid blister test} is on the order of 5 mm, while the product $L(\gamma/Eh)^{1/4}$ ranges between 1 and 5 mm.\\

From a quantitative point of view, the prefactor in equation (\ref{eqelcap}) can be found by using an exact expression for the the elastic energy of the blister, or equivalently by using the fact that the curvature at the adhesion point (and at the crest) is 
 equal to $2/L_{ec}$ \cite{review10}. Equation (\ref{eqelcap}) becomes finally \cite{vella2009}: 
 \begin{equation}
 \lambda^2/d=\pi^2  L_{ec}.
\label{eqLecpref}
 \end{equation}
However, fitting the experimental data (figure~\ref{fig3}) results in a slightly different prefactor $\lambda^2/d=7.3  L_{ec}$. 
This mismatch is not due to an effect of the stretching strain. Indeed the gaussian curvature of the crest of the blisters appears larger than in the case of a collection of independant one-dimensional blisters, which leads to an higher stretching energy.
The origin of the discrepancy certainly lies in the experimental preparation of the blister. 
During the evaporation of the wetting liquid layer, temporary large depression leads to smaller, more curved blisters. 
This might lead to some plastic deformation in the material (plasticity is expected in polypropylene for a radius of curvature smaller than 100$h$). 
For the 15$\mu\,$m sheets, the curvature expected in the blister is 3mm, which is not very far from 100$h=1.5\,$mm. However no clear signs of plasticity have been observed in the experiments.

\subsection{Thin sheets on soft substrates under compression}

In this paper we focus on the situation of an adhesive sheet deposited on a rigid passive substrate, and blisters
appear because of imposed lateral displacements (controlled, as in section 2, or randomly in section 3).
But another  common situation is found when thin films deposited on a substrate delaminate under compressive stresses induced by the deposition process, thermal expansion, chemical swelling, or mechanical compression of the substrate. For instance, a recent study motivated by potential applications to stretchable electronics focused on the formation of one-dimensional blisters observed when a soft substrate covered with a thin film is  uni-axially compressed \cite{vella2009}. 
Prior to delamination, wrinkles are commonly observed as the result of the balance between the bending energy of the film and the stretching energy of the substrate \cite{bowden98}. The wavelength of such wrinkles scales as $\lambda \sim h(E/E_s)^{1/3}$, where $E$ and $E_S$ are the Young moduli of the film and the substrate, respectively. 
However we focus here on delamination where the film detaches from the substrate.
In this case, three physical ingredients dictate the buckling behavior of the thin layer: adhesion, bending stiffness and, in addition, the compression of the substrate. 
Although the condition to delaminate relies on a combination of these three ingredients, this study shows that the shape of one-dimensional blisters is set by the balance between adhesion and bending rigidity, identical to the case of a thin sheet able to slide on a rigid plate (the transverse displacement depends however on the Young's modulus of the substrate).

\begin{figure}[!h]
\centering
\includegraphics[width=8cm]{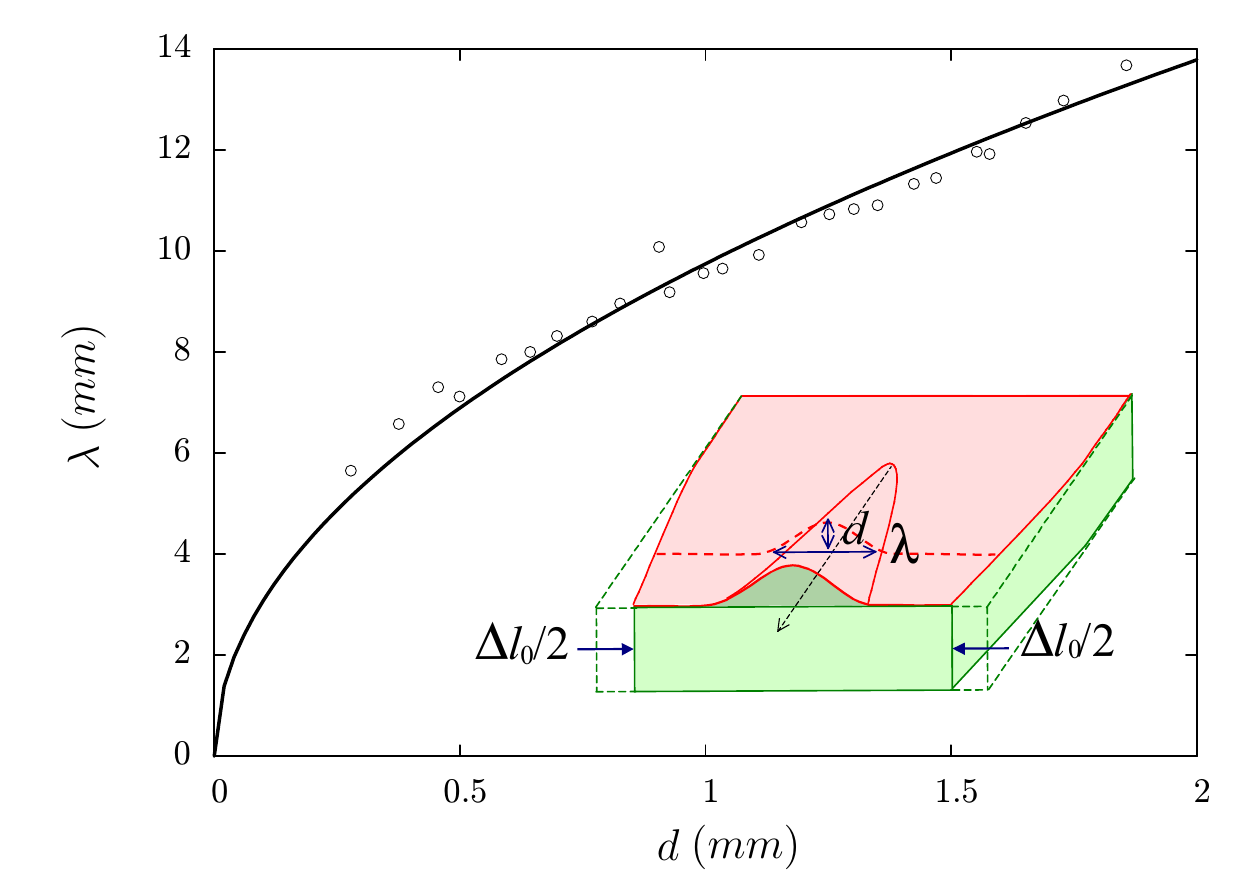}
\caption{Compression of a soft substrate covered with a thin film (insert: sketch of the experiment): as in the rigid substrate configuration, the curvature of the crest is found constant ($\lambda \sim L_{ec}^{1/2}d^{1/2}$) as prescribed by Equation~\ref{eqLecpref}.}
\label{fig9}
\end{figure}

We assume here that the possible formation of wrinkles has a negligible impact while the width of the blisters is large in comparison with the wavelength of these wrinkles.
We thefore expect that our results on angular blisters (eqn.~\ref{eqelcap}) extend to the case of the delamination of thin films from soft substrates if condition eq.~\ref{eqnostretch} is satisfied. 
To validate this assumption, a sheet of polypropylene ($E=2.1\pm0.2\,$GPa, $h = 30\,\mu$m) is uniformly laid on a block of  thickness $h_S = 5\,$cm made with vinylpolysiloxane (Zhermack) of Young's modulus $E_S \simeq 220\,$kPa. The thin film adheres to the substrate through Van der Waals interactions.
The block is  obliquely compressed with non-parallel clamps, which induces the propagation of one or multiple angular blisters in the direction perpendicular to the compressive load (sketch in the insert of figure~\ref{fig9}). 
The shape of a typical blister reported in figure~\ref{fig9} follows the same form $\lambda \sim L_{ec}^{1/2}d^{1/2}$ as in the rigid substrate configuration. The length of the blister $L \simeq 80\,$mm is sufficient to fulfill the condition (5).
The deduced value of the \textit{elasto-capillary} length $L_{ec} = 9.6\,$mm is in good agreement with the estimation from material properties and with the prefactor for the law $\lambda \sim L_{ec}^{1/2}d^{1/2}$ found in Vella \textit{et. al.}\cite{vella2009}.\\

In conclusion to the simplified case of a single angular blister with a small slope, the radius of curvature of the crest is found constant and proportional to $L_{ec}$. Measuring this radius of curvature thus 
allows to estimate the \textit{elasto-capillary} length, a typical length scale which compares bending stiffness and adhesion, (or one of these quantities if the other one is known). The 
robustness of this result to simultaneous random blisters is probed in the following section.

\section{Multiple random blisters}
We present in this section the generalized case of multiple blisters obtained as a thin sheet is randomly laid on an adhesive substrate, and also discuss the shape of the tips of the blisters.
 
Contrary to the model experiment with a single angular blister, the random blisters may be closed and trap a certain volume of air between the thin sheet and the substrate. We assume here that the pressure is small enough and do not influence the shape of the blister$^4$.\footnote{$^4$ The effect of a pressure $p$ is negligible if the associated energy per unit surface $  p d$ is 
much smaller than  $\gamma $, or equivalently $p\ll\gamma/d$, where $d$ is the heigh of the blister.} In the case of high pressure, the contact line would tend to be circular \cite{dannenberg61, williams69}, in contrast with the elongated shapes observed in our experiments.

\subsection{Curvature profiles}
\begin{figure}[!h]
\begin{minipage}[b]{0.5\textwidth}
\subfigure[]{\includegraphics[width=7.5cm]{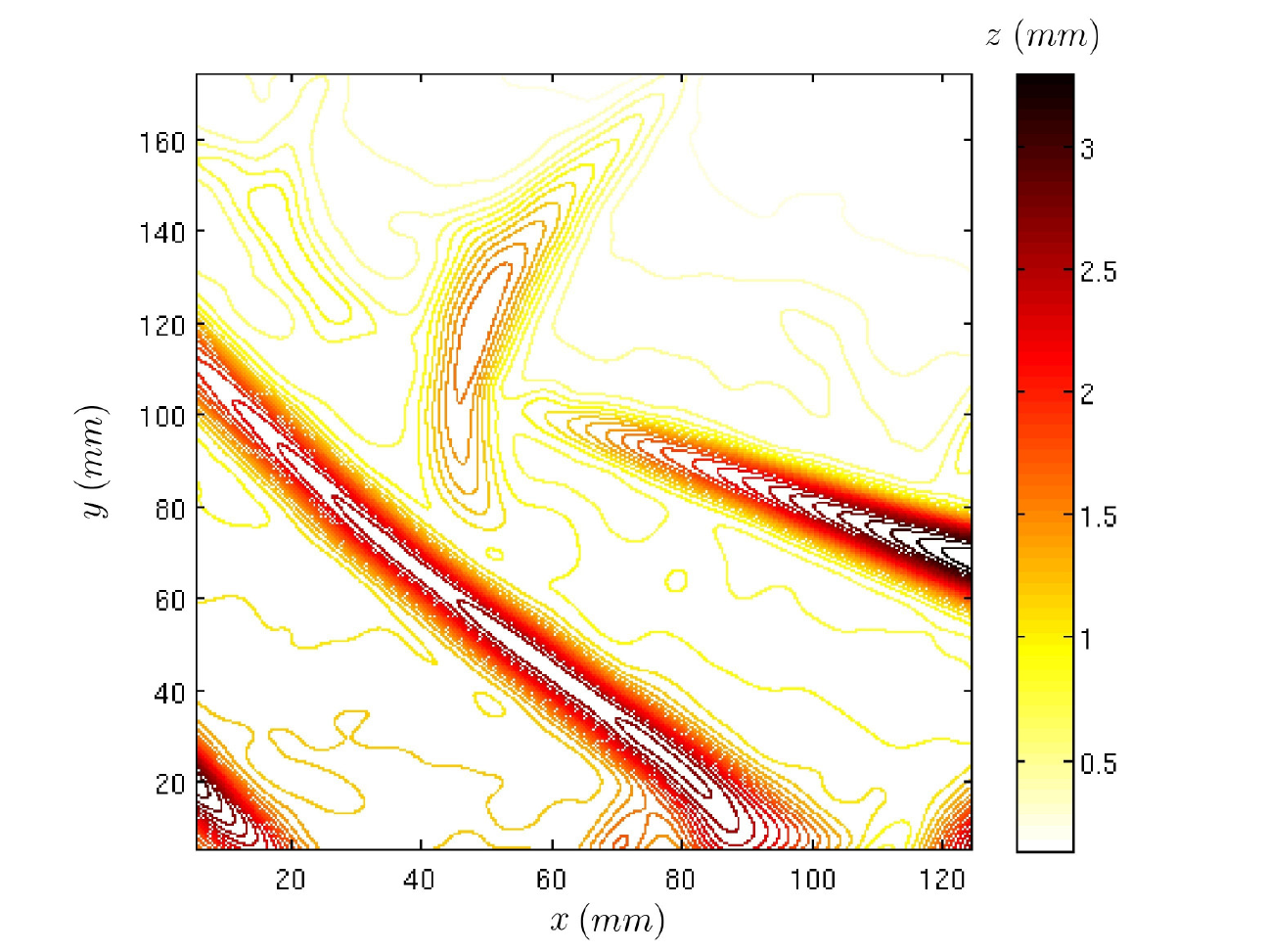}}
\end{minipage}
\begin{minipage}[b]{0.5\textwidth}
\subfigure[]{\includegraphics[width=7.5cm]{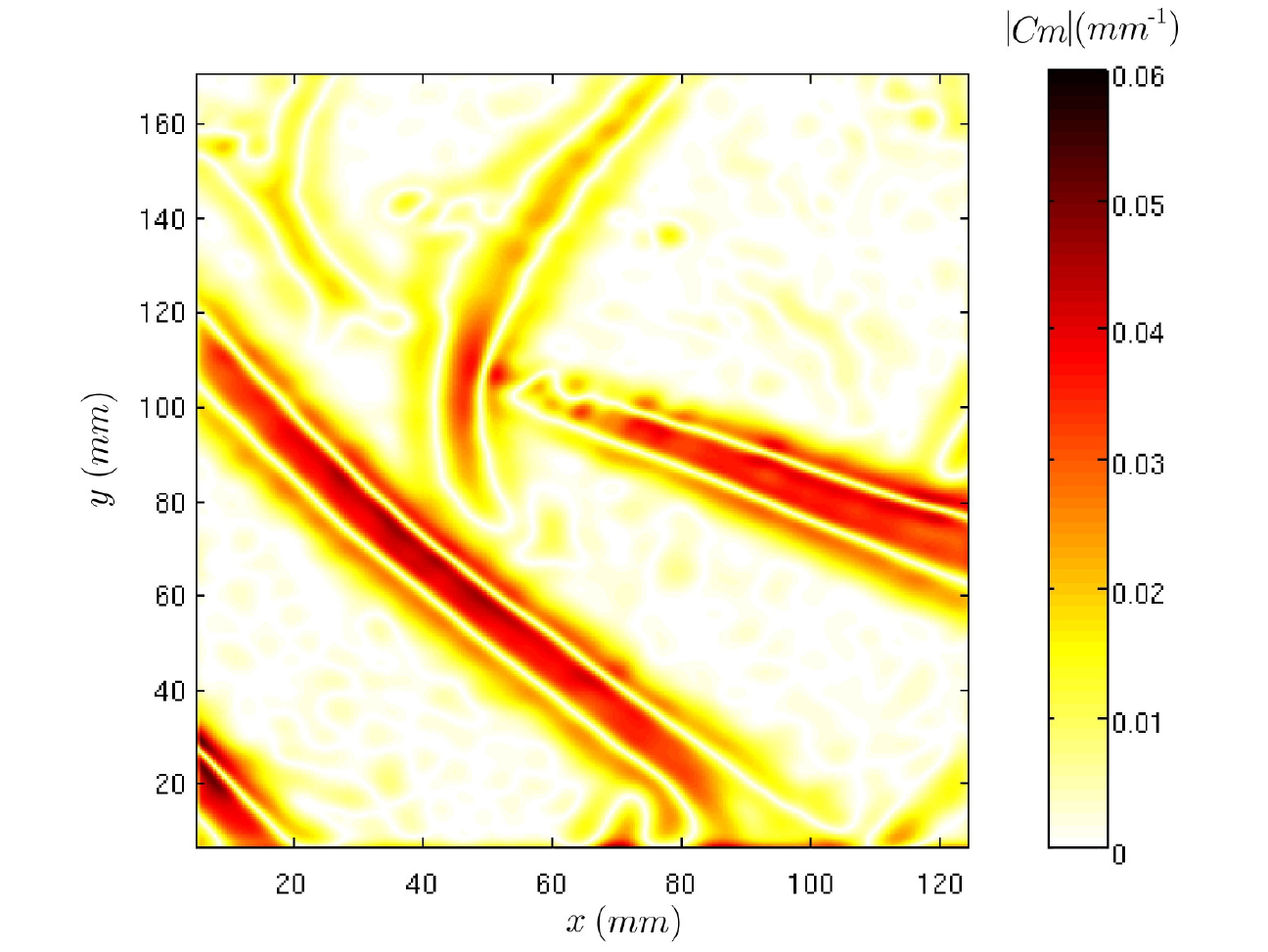}}
\end{minipage} 
\begin{center}
\subfigure[]{\includegraphics[width=4cm]{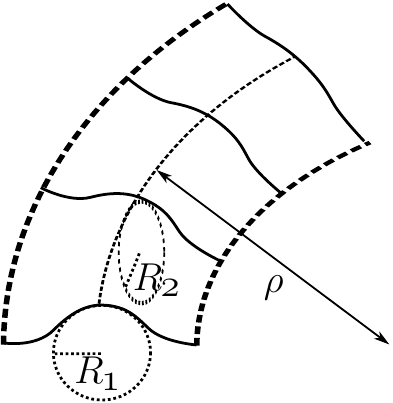}}
\end{center}
\caption{(a) Height profile of random blisters formed when a thin sheet is carelessly deposited over an adhesive plate. (b) Field of absolute mean curvature $|C_m|=|1/2(1/R_1+1/R_2)|$, where $R_1$ and $R_2$ are the principal radii of curvature. (c) Sketch of the different radii of curvature.}
\label{fig5}
\end{figure}

A rigid plate is preliminary coated with a thin layer of vinylpolysiloxane, which provides adhesion through van der Waals interactions. The \textit{elasto-capillary} length is preliminary measured from a one-dimensional blister and gives $L_{ec} \simeq 22\,$mm. Multiple blisters are randomly formed afterward by depositing carelessly a thin sheet of polypropylene ($E=2.1\pm0.2\,$GPa, $h = 30\,\mu$m) over the plate.
The height profile of the whole sheet is determined with a versatile optical profilometry technique based on the deformation of a fringe pattern projected on the surface \cite{cobelli2009, maurel2009}.
The field of mean curvature $C_m=1/2(1/R_1+1/R_2)$, where $R_1$ and $R_2$ are the principal radii of curvature, is readily deduced from the topography with a standard MatLab routine. 
Although blisters with different heights and slopes are observed (figure~\ref{fig5}a), the curvature of the crests remains almost uniform for the different blisters (figure~\ref{fig5}b).
In the particular case illustrated in figure~\ref{fig5}, we find $C_m \sim 0.04\,\mathrm{mm}^{-1}$, which corresponds to $L_{ec} = 1/C_m \simeq 25\,$mm in good agreement with a preliminary measurement with a one-dimensional blister.

One blister in the middle of the figure~\ref{fig5}b
seems however significantly different from the others: its crest is not straight but slightly curved, with a typical radius of curvature $\rho$ sketched in figure \ref{fig5}c. As friction prevents re-arrangements of the adhering patches, such curved blisters tend to appear randomly depending on the way the film has been deposited.  This curvature induces gaussian curvature, and therefore an additional source of stretching energy. An estimate of  this gaussian curvature $K$  is given by the product of the typical curvature of a section of the blister (\textit{i.e.} $1/L_{ec}$, as shown in the previous section) by the curvature of its axis projected on the normal to the sheet (inclined with a typical angle $d/\lambda$) \cite{struik88}: $K \sim (1/ L_{ec})(d/\lambda \rho)$.  To compute the typical strain $\epsilon$, we use the fact that Gauss'\textit{Theorema Egregium} can be written as $K \sim \partial^2\epsilon$ \cite{pauchard97, witten07}. This scaling can also be deduced from the second F\"oppl-von K\'arm\'an compatibility equation $\Delta^2 \phi + E K = 0$ \cite{landau}, where $\phi$ is the Airy stress function (with $\partial^2 \phi \sim E\epsilon$) and $E$ the Young modulus of the material. Since the width $\lambda$ corresponds to the typical length of variation of the strain in the blister, the strain can be estimated as $\epsilon \sim \lambda d/\rho L_{ec}$. Note that in the case of straight angular blisters, the same derivation leads to $K \sim d^2/\lambda^2 x^2$, and thus to an estimate of the strain $\epsilon \sim d^2/x^2$, which is consistent with the equation \ref{eqnostretch}.
The density of stretching energy of a curved blister is thus on the order of $Eh\lambda^2 d^2/\rho^2 L_{ec}^2$. Neglecting the effect of the curvature of the blister is only valid if $Eh\lambda^2 d^2/\rho^2 L_{ec}^2 \ll \gamma$, {\it i.e.} $\rho \gg \lambda d/h$. For the experiment shown in figure~\ref{fig5}, the critical radius beyond which this additional stretching energy can be neglected is on the order of $700\,$mm, which is higher  than the typical radius observed for this particular blister ($\rho\simeq 50\,$mm). The additional stretching energy should explain the peculiar profile of this curved blister.

Note finally that in the model experiment described in the previous section, the movement of the sides of the single blister is a rotation, whose center is located at the tip of the blister.
In the case of multiple random blisters, each tip of a blister is a local center of rotation of the adhered parts of the thin sheet.

It would be tempting to extrapolate our macroscopic results down to the molecular scale of graphene sheets. 
For instance, Xu \textit{et al.} \cite{xu09} recently reported blisters on monolayer graphene sheets deposited on $\mathrm{SiO}_2$ substrates of widths, heights and lengths on the order of  $10\,$nm, $3\,$nm and $1\,\mu$m, respectively. 
We can thus estimate from the ratio $\lambda^2/d$ the corresponding \textit{elasto-capillary} length $L_{ec}\sim 3\,$nm. 
Assuming a bending stiffness for a monolayer on the order of $0.2\,\mathrm{nN}.\mathrm{nm}$ derived from phonon spectrum of graphite, potentials or \textit{ab initio} calculations \cite{patra09, lu2009, cranford09}, we find an adhesion energy$^5$ $\Delta\gamma$ on the order of  $50\,\mathrm{mJ.m}^{-2}$.
\footnote{$^5$ When adhesion is provided by a wetting liquid, the work of adhesion is equal to $2\gamma$. In the case of conventional adhesion, the adhesion energy $\Delta \gamma$ is thus equivalent to $2\gamma$ in the definition of $L_{ec}$.}
Although {\it ab initio} calculations predict an adhesion energy one order of magnitude higher \cite{li2009}, large mismatches have been reported between calculated and measured values in similar situations \cite{zong10}. Moreover, the height of these blisters are on the order of the typical decay length of Van der Waals interactions \cite{li2009} and such interactions cannot be neglected to compute the shape of the  buckled part of the graphene sheet. Progresses have still to be made to produce a reliable estimate of the adhesion energy at those microscopic scales, but the observation of blisters may constitute a valuable method.

\subsection{Radius of the tip}
We finally focus on another feature of these small amplitudes blisters: the shape of the tip. The scalings previously shown for the width and amplitude of blisters (equation \ref{eq})  are no longer valid near the tip, where they would lead to a diverging curvature of the crest line.
Indeed,  since $d(x)/x$ is proportional to $ x^{-1/3}$, the local version of condition (\ref{eqnostretch}) is always violated close to the tip of the blister. As a matter of fact, any smooth regularization of the tip  necessarily involves non-zero gaussian curvature, producing in-plane deformations. We thus assume that all three energies (bending, stretching, and adhesion) are of same order in this region.
If the tip ends with a circular contact line (radius $R)$, the stretching energy density involved in the spherical cap of amplitude $d$ is proportional to $Eh (d/R)^4$ following an estimate similar to equation~\ref{eqnostretch}, while the bending energy density scales as $Eh^3 d^2/R^4.$ Imposing that both energies are comparable to the adhesion energy $\gamma$ leads  
 to  $d\sim h$ and $R^2/d \sim L_{ec}$. We finally find that the radius of the tip is given by:
\begin{equation}
R \sim \sqrt{hL_{ec}}.
\label{rtip}
\end{equation}
 
\begin{figure}[!h]
\centering
\includegraphics[width=8cm]{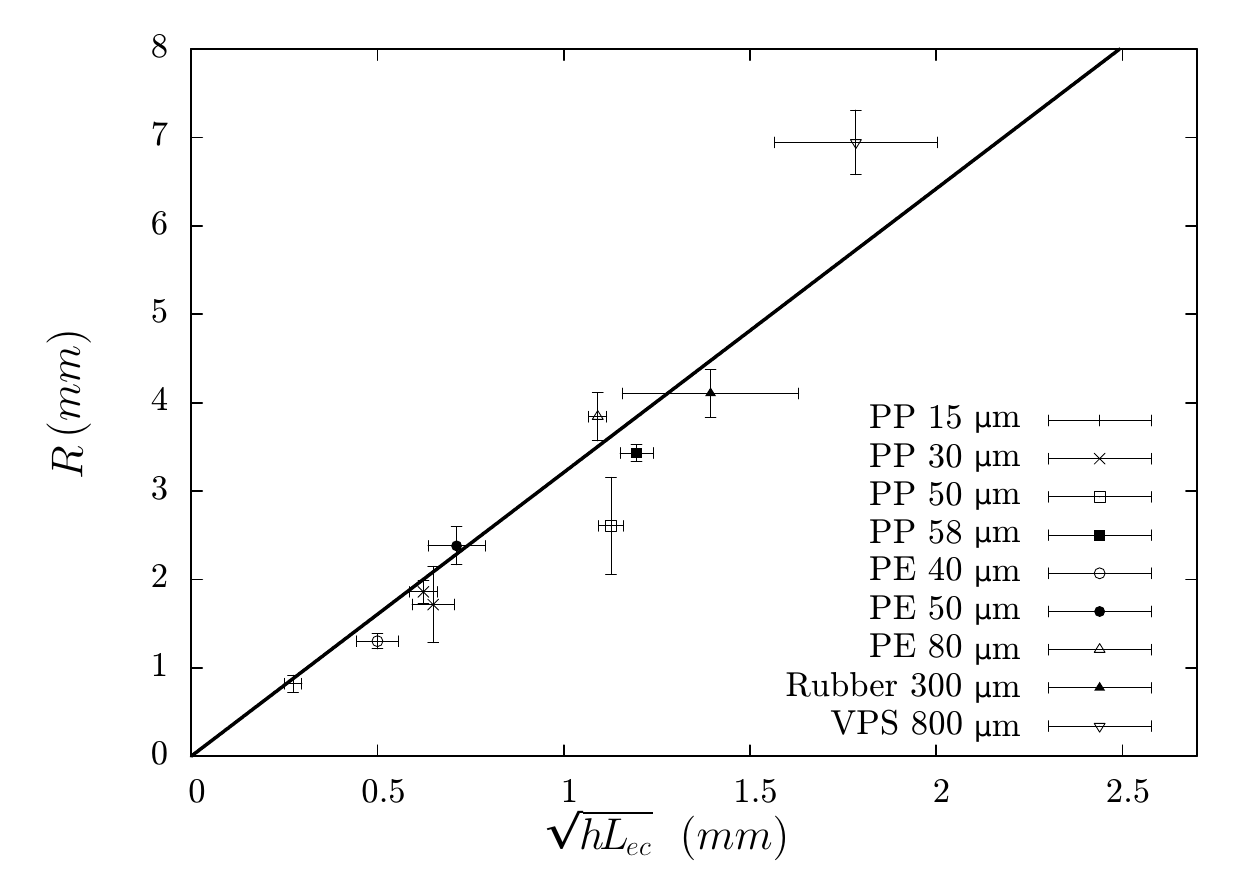}
\caption{Radius of the tip $R$ as a function of $\sqrt{hL_{ec}}$ for sheets of various polymeric materials and thicknesses ranging from $15\,\mu$m to $1\,$mm (Equation (\ref{rtip})). $R$ can be approximated by $R = 3.2\sqrt{hL_{ec}}$.}
\label{fig7}
\end{figure}

A more quantitative description of the shape of the tip would require solving F\"oppl-von K\'arm\'an equations numerically. We nevertheless propose here a simplified model where we assume that the shape of the tip is the half of an axisymetric cap of radius $R$ and height $d$. The Lagrangian \cite{timoshenkoplate} (or more commonly, the energy) of the cap can be approximated by$^6$\footnote{$^6$ See paragraph 100 in \cite{timoshenkoplate}, valid for deflections of order $h$.}:
\begin{equation}
\mathcal{L} = \frac{4\pi B d^2}{R^2} + \frac{\alpha \pi E h d^4}{R^2} + \pi R^2 \gamma +  \beta(d-d_{tip}),
\end{equation}
where $\alpha$ is a numerical constant ($\alpha \approx 0.07$) and $\beta$ the Lagrange multiplier that imposes the continuity of height (at $d=d_{tip}$) between the tip and the remaining part of the blister. Minimizing $\mathcal{L}$ with respect to $R$, $d$ and $\beta$ yields:
\begin{equation}
R^4 = \frac{4Bd_{tip}^2 + \alpha E h d_{tip}^4}{\gamma}. 
\end{equation}
The amplitude $d_{tip}$ has to be related with the height beyond which an axisymetric blister is no longer stable, so, as pointed out in \cite{chopin2008}, $d_{tip} \approx \delta h$, where $\delta$ is a numerical prefactor:
\begin{equation}
R = (4\delta^2 + 12\alpha(1-\nu^2)\delta^4)^{1/4}\sqrt{hL_{ec}}. 
\end{equation}
We find here the prefactor for the proposed scaling law (equation (\ref{rtip})), but it depends on the additional parameter $\delta$.
Experimental data obtained using films of various polymeric materials show that the radius of the tip varies linearly with $\sqrt{hL_{ec}}$ and can be approximated by $R = 3.2\sqrt{hL_{ec}}$ (figure~\ref{fig7}). Despite the variations on both the measurements of $R$ and the \textit{elasto-capillary} length$^7$, as shown by the error bars in figure~\ref{fig7}, this result appears significant knowing that the thicknesses used range from $15\,\mu$m to $1\,$mm. 
\footnote{$^7$ The value for adhesion energy of the polymer on the vinylpolysiloxane  depends if the measurement corresponds to sticking a lamella of polymer on the substrate (lower adhesion), or to debounding (higher adhesion). The horizontal error bars in figure~\ref{fig7} have been determined with these extremes values, obtained by measuring the shape of one-dimensional blisters.}
The prefactor found experimentally yields an estimate of the parameter $\delta \simeq 3.1$. Despite the fact Chopin \textit{et al.} noticed that the threeshold for the instability of the axisymetric blister was not reproducible in their experimental set-up, we can estimate from their data a critical amplitude $d_c \approx 50 h$. However, in the {\it liquid blister case}, the shape loses its symmetry and becomes triangular after the onset of the instability.  
Further work is still necessary to characterize completely the shape of the tips of these delamination blisters.

We conclude that while the ratio $\lambda^2/d$ provides an estimate of $L_{ec}$, measuring $R$ gives an additional information on the thickness of the sheet. Simple topographic characterization of random blisters finally allow to access both relevant length scales of the problem.

\section{Conclusion}

Although the formation of blisters on thin sheets deposited on a substrate is usually viewed as an undesirable defect, a simple geometrical characterization of low amplitude blisters provides estimates of both the thickness of the sheet and an {\it elasto-capillary} length scale $L_{ec}=\sqrt{B/\gamma} $ that compares bending stiffness $B$ and adhesion $\gamma$. The radius of curvature of the crest is found uniform through the whole sample and proportional to $L_{ec}$ while the radius of the tip is proportional to $\sqrt{hL_{ec}}$. Scaling down our macroscopic experiments may allow to estimate the mechanical properties of thin polymeric films or ultimately of molecular films such as graphene sheets that are difficult to assess.

\vspace{1cm}

\noindent\textbf{Acknowledgements}

\noindent
We thank Pablo Cobelli for allowing and helping us to use his profilometer technique and Ko 
Okumura for fruitful discussions.
Y. A. was supported by a joint scholarship between the French government and
Ochanomizu University (ITP program from the Japan Society for the Promotion 
of Science).
This work was partly funded by a French ANR junior project (MecaWet).

\end{document}